\documentclass[prl,twocolumn,superscriptaddress,showpacs,floatfix]{revtex4}
\usepackage{graphicx,amsmath,amssymb,bm}
\usepackage{amsfonts}

\begin{document}

\title{Evolution of shell structure in neutron-rich calcium isotopes}

\author{G.~Hagen}
\affiliation{Physics Division, Oak Ridge National Laboratory,
Oak Ridge, TN 37831, USA}
\affiliation{Department of Physics and Astronomy, University of
Tennessee, Knoxville, TN 37996, USA}
\author{M.~Hjorth-Jensen}
\affiliation{Department of Physics and Center of Mathematics for Applications, University of Oslo, N-0316 Oslo, Norway}
\affiliation{National Superconducting Cyclotron Laboratory and Department of Physics and Astronomy, Michigan
  State University, East Lansing, MI 48824, USA}
\author{G.~R.~Jansen}
\affiliation{Department of Physics and Center of Mathematics for Applications, University of Oslo, N-0316 Oslo, Norway}
\author{R.~Machleidt}
\affiliation{Department of Physics, University of Idaho, Moscow, ID 83844, USA}
\author{T.~Papenbrock}
\affiliation{Department of Physics and Astronomy, University of
Tennessee, Knoxville, TN 37996, USA}
\affiliation{Physics Division, Oak Ridge National Laboratory,
Oak Ridge, TN 37831, USA}

\begin{abstract}
  We employ interactions from chiral effective field theory and
  compute the binding energies and low-lying excitations of calcium
  isotopes with the coupled-cluster method.  Effects of three-nucleon
  forces are included phenomenologically as in-medium two-nucleon
  interactions, and the coupling to the particle continuum is taken
  into account using a Berggren basis.  The computed ground-state
  energies and the low-lying $J^\pi=2^+$ states for the isotopes
  $^{42,48,50,52}$Ca are in good agreement with data, and we predict
  the excitation energy of the first $J^\pi=2^+$ state in $^{54}$Ca at
  $1.9$~MeV, displaying only a weak sub-shell closure.  In the
  odd-mass nuclei $^{53,55,61}$Ca we find that the positive parity
  states deviate strongly from the naive shell model.
\end{abstract}


\maketitle

{\it Introduction.} -- The shell model is the paradigm for our
understanding of atomic nuclei. Doubly magic nuclei (i.e., nuclei that
exhibit an enhanced stability) are its cornerstones, and they
determine the properties of nuclei in entire regions of the nuclear
chart. The magic numbers -- established {\it ad hoc} via a mean field
plus a strong spin-orbit interaction more than 60 years ago by Mayer
and Jensen for beta-stable nuclei~\cite{MJ} -- are modified in
neutron-rich nuclei, see for example Ref.~\cite{sorlin2008} for a
recent review.  The magic nature of nuclei is reflected experimentally
in enhanced neutron separation energies and a reduced quadrupole
collectivity (i.e., a relatively high-lying first excited
$J^{\pi}=2^+_1$ state and relatively small electromagnetic transition
probabilities from this state to the $J^{\pi}=0^+$ ground state).  In
doubly magic nuclei such as $^{40}$Ca and $^{48}$Ca, the $J^{\pi}=2^+$
state appears at an excitation energy close to $4$~MeV, while in
open-shell calcium isotopes like $^{42,44,46,50}$Ca, this excitation
energy is closer to $1$~MeV. How these quantities
evolve as we move towards the driplines is an open issue in ongoing
nuclear structure research and is intimately related to our
fundamental understanding of shell evolution in nuclei.

For the theoretical understanding of shell evolution, phenomenological
terms such as the tensor interaction~\cite{otsuka2005} have been
proposed.  In a modern picture, three-nucleon forces (3NFs) play a
pivotal role in shell evolution~\cite{zuker2003}. In the oxygen
isotopes, for instance, 3NFs make $^{24}$O doubly magic and a dripline
nucleus~\cite{otsuka2010,jdholt2011b,hagen2012}.  Similarly, Holt {\it
  et al.}~\cite{jdholt2011} showed that 3NFs are greatly responsible
for the magic neutron number $N=28$. On the other hand, experiment and
theory show that the next possible magic number, $N=32$, exhibits a
smaller value for the $2^+$ excitation (but more than twice as large
as seen in open-shell calcium isotopes) than observed in $^{48}$Ca.
This is often referred to as a sub-shell closure.  The $N=32$
sub-shell closure is well established from experiments in
calcium~\cite{huck1985,gade2006}, titanium~\cite{janssens2002}, and
chromium~\cite{prisciandaro2001}. However, the situation is more
complicated for neutron-rich calcium isotopes.  For the neutron number
$N=34$, no sub-shell closure is seen experimentally in
chromium~\cite{marginean2006} or
titanium~\cite{liddick2004,dinca2005}, and there are some doubts
regarding a sub-shell closure in calcium~\cite{rejmund2007}. Different
theoretical predictions have been made around $N=34$. Within the $fp$
shell-model space, the empirical interaction GXPF1~\cite{gxpf1}
predicts a strong shell gap in $^{54}$Ca, while the monopole-corrected
KB3 interaction~\cite{kb3} yields no shell gap. A low-momentum
shell-model interaction with empirical single-particle energies and a
$^{48}$Ca core yields a weak sub-shell closure in $^{54}$Ca~\cite{luigi2009}.
Shell-model calculations that include 3NFs predict a shell closure in
$^{54}$Ca in the $fp$ model space, and this shell closure is reduced
to a sub-shell closure (similar in strength to the $N=32$ sub-shell
closure in $^{52}$Ca) in an enlarged model space that also includes
the $g_{9/2}$ orbital~\cite{jdholt2011}. Thus, the picture regarding
the shell gap in $^{54}$Ca is not settled yet. The theoretical
prediction of the shell evolution in calcium isotopes is a challenging
task that requires a very good understanding of the nuclear
interaction, accurate treatment of many-body correlations and coupling
to the scattering continuum~\cite{jacek2007}. To study the shell
evolution, we will focus on neutron separation energies, the energies
of the first excited $J^\pi=2^+$ states, and spectra in the nuclei
$^{53,55,61}$Ca, which differ by one neutron from nuclei that exhibit
a closed subshell in the naive shell model.

In this Letter, we present a state-of-the-art prediction for the shell
evolution of neutron-rich calcium isotopes. To this purpose, we employ
nucleon-nucleon ($NN$) interactions from chiral effective field theory
(EFT) together with a schematic approximation of 3NFs guided by chiral
EFT, and utilize the coupled-cluster method to solve the quantum
many-body problem. Chiral EFT is a systematic and model-independent
approach to nuclear interactions. We employ the $NN$ interactions at
next-to-next-to-next-to leading order by Entem and
Machleidt~\cite{entem2003, machleidt2011}, and an approximation for
the chiral 3NFs that was previously adopted in neutron-rich oxygen
isotopes~\cite{hagen2012}. The coupled-cluster
method~\cite{ccm,bartlett07} is a very efficient tool for the
computation of nuclei with a closed (sub-)shell structure and their
neighbors, and thus ideally suited for the task at hand.

{\it Hamiltonian, model space, and method.} -- We employ the intrinsic
Hamiltonian
\begin{equation}
\label{ham}
\hat{H} = \sum_{1\le i<j\le A}\left({(\vec{p}_i-\vec{p}_j)^2\over 2mA} + \hat{V}
_{NN}^{(i,j)} +  \hat{V}_{\rm 3N eff}^{(i,j)}\right) \ .
\end{equation}
Here, the intrinsic kinetic energy depends on the mass number $A$.
The potential $\hat{V}_{NN}$ denotes the chiral $NN$ interaction at
next-to-next-to-next-to leading order~\cite{entem2003,machleidt2011},
while $\hat{V}_{\rm 3N eff}$ is a schematic potential based on the
in-medium chiral $NN$ interaction by Holt~{\it et al.}~\cite{holt2009}
(see also Ref.~\cite{hebeler2011}). The potential $\hat{V}_{\rm 3N
  eff}$ results from integrating one nucleon in the leading-order
chiral 3NF over the Fermi sphere with Fermi momentum $k_F$ in
symmetric nuclear matter and is thus reminiscent of the normal-ordered
approximation~\cite{hagen2007}. It depends formally on the Fermi
momentum $k_F$, the low-energy constants $c_D$ and $c_E$ of the
short-ranged contributions to the leading-order chiral 3NF, and the
chiral cutoff. The latter is equal to the value employed in the chiral
$NN$ interaction~\cite{entem2003}. In the computation of neutron-rich
oxygen isotopes~\cite{hagen2012}, the parameters $k_F=1.05$~fm$^{-1}$
and $c_E=0.71$ resulted from adjusting the binding energies of
$^{16,22}$O, while $c_D=-0.2$ was kept at its value determined in
light nuclei~\cite{gazit}. In this work, we use $k_F=0.95$~fm$^{-1}$
and $c_E=0.735$ from adjusting the binding energies of $^{48,52}$Ca.
It is very satisfying that the parameterization of $\hat{V}_{\rm 3N
  eff}$ changes only little as one goes from neutron-rich isotopes of
oxygen to the significantly heavier calcium isotopes.

The coupled-cluster method generates a similarity-transformed
Hamiltonian $\overline{H}=e^{-T}\hat{H}e^T$ by the action of the
cluster operator $T$ that creates up to $n$-particle--$n$-hole
($np$-$nh$) excitations with respect to a reference state. Details of
our implementation are presented in Refs.~\cite{hagen2008,hagen2010a}.
We compute the ground states of the closed-(sub)shell nuclei
$^{40,48,52,54,60,62}$Ca in the singles doubles (CCSD) approximation
and include $n=3$ triples perturbatively within the $\Lambda$-CCSD(T)
approach of Ref.~\cite{taube2008}. Our model space consists of up to
$N_{\rm max}=19$ major spherical oscillator shells (i.e., the maximal
single-particle excitation energy is 18 units of $\hbar\omega$ above
the oscillator ground state), and the reference state results from a
Hartree-Fock calculation. Our basis employs oscillator spacings
between $24\mbox{~MeV}\le\hbar\omega\le 32\mbox{~MeV}$, and in the
largest model spaces the results we present are practically
independent of $\hbar\omega$. For excited states in $^{53,55,61}$Ca
above threshold, we use a Gamow-Hartree-Fock
basis~\cite{michel,hagen2012} with 40 discretization
points~\cite{hagen2006}.  Excited states of the closed-shell nuclei
$^{48,52,54}$Ca are computed within the equation-of-motion (EOM) method
with singles and doubles.  The open-shell nuclei
$^{39,41,47,49,51,53,55,59,61}$Ca are computed with the particle
attached/removed EOM methods, and we use the two-particle attached EOM
method~\cite{jansen2011} for the nuclei $^{42,50,56}$Ca.  Note that
the employed EOM methods are expected to reliably compute the
separation energies and low-lying excited states as long as they are
dominated by 1$p$, 1$h$, 1$p$-1$h$ or 2$p$ excitations.

\begin{figure}[htbp]
  \begin{center}
    \includegraphics[width=0.45\textwidth,clip=]{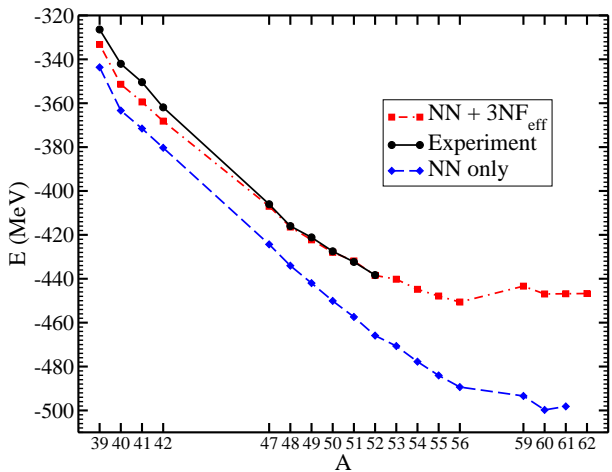}
  \end{center}
  \caption{(Color online) Ground-state energy of the calcium isotopes
    as a function of the mass number $A$. Black circles:
    experimental data; red squares: theoretical results including the effects of
    three-nucleon forces; blue diamonds: predictions from chiral $NN$ forces alone. 
    The experimental results for $^{51,52}$Ca are from Ref.~\cite{Dilling}.}
  \label{fig1}
\end{figure}

{\it Results.} -- Figure~\ref{fig1} shows the computed ground-state
energies of the calcium isotopes and compares the results obtained
with the Hamiltonian of Eq.~(\ref{ham}) to available data and to the
results based on chiral $NN$ interactions alone. The inclusion of
chiral 3NFs via the in-medium effective potential $\hat{V}_{\rm 3N
  eff}$ clearly yields a much improved agreement with data. The light
isotopes $^{39,40,41,42}$Ca are slightly overbound, while the
agreement is very good for the neutron-rich isotopes at the center of
this study. The comparison with chiral $NN$ forces shows that the
in-medium effective potential $\hat{V}_{\rm 3N eff}$ is
repulsive~\footnote{The binding energy of $^{40,48}$Ca resulting from
  chiral $NN$ interactions alone is larger than previously reported in
  Refs.~\cite{hagen2008,hagen2010a}. In the present work we use a
  higher cutoff ($J_{\rm max}=10$) in the relative angular momentum
  when transforming the $NN$ interaction from the center-of-mass frame
  to the laboratory frame.}. For the heavier isotopes of calcium, the
in-medium effective potential $\hat{V}_{\rm 3N eff}$ becomes
increasingly repulsive, and a saturation of the total binding energy
sets in around $^{60}$Ca.  It is interesting that essentially the same
interaction yields attraction in neutron-rich oxygen
isotopes~\cite{hagen2012}.  Note that the results for the isotopes
$^{52-60}$Ca are based on an exponential extrapolation of our results
for $N_{\rm max}=14,16,18$ oscillator shells at the oscillator
frequency $\hbar\omega=26$~MeV.  This
extrapolation adds 0.6~MeV of binding energy in $^{52}$Ca and 1.2~MeV
in $^{60}$Ca. For the other nuclei our results are practically
converged with $19$ shells. As a check, we also employed a Gamow
Hartree-Fock basis for the $s$-$p$-$d$ neutron partial waves and
computed the ground states for the isotopes $^{54,60}$Ca. The
differences are small (up to about 0.3~MeV) and barely noticeable on
the scale of Fig.~\ref{fig1}.

\begin{figure}[htbp]
  \begin{center}
    \includegraphics[width=0.45\textwidth,clip=]{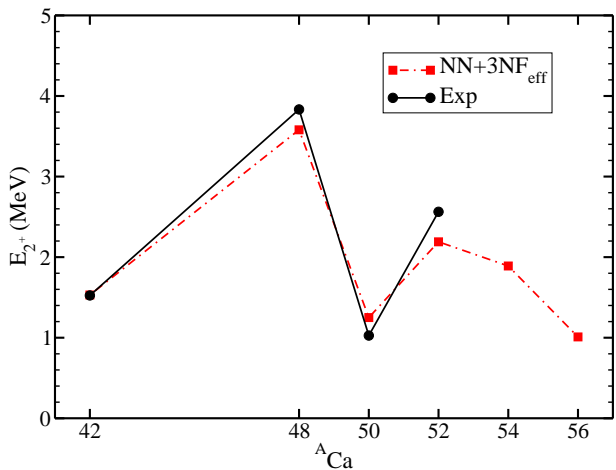}
  \end{center}
  \caption{(Color online) Excitation energies of $J^\pi=2^+$ states in
    the isotopes $^{42,48,50,52,54,56}$Ca (experiment: black, theory:
    red).}
  \label{fig2}
\end{figure}

Figure~\ref{fig2} shows the energies of the $J^\pi=2^+$ states in the
isotopes $^{42,48,50,52,54,56}$Ca, computed in a model space with
$N_{\rm max}=18$ and $\hbar\omega=26$~MeV for $^{48,52,54}$Ca and
$N_{\rm max}=16$ and $\hbar\omega=28$~MeV for $^{42,50,56}$Ca.  Where
data is available, we obtain a good agreement between theory and
experiment. The 3NFs generate the shell closure in $^{48}$Ca and make
$N=28$ a magic number. We also computed the $J^\pi=2^+$ excited state
in $^{48}$Ca using chiral $NN$ forces alone, and we obtained $E_{2^+}=
2.07$ MeV; this shows that the magicity of $^{48}$Ca is due to the
effects of 3NFs in our approach.  Our results predict that $^{54}$Ca
exhibits only a weak sub-shell closure, and this is one of the main
results of this Letter.

To further demonstrate this point, we also computed the
energies of the first excited $J^\pi=4^+$ states, and neutron
separation energies $S_n$. The results in Table~\ref{tab1} show that
theory and experiment agree well for the available data. The
theoretical results for $^{54}$Ca suggest strongly that this nucleus
will only exhibit a weak sub-shell closure but not a shell closure. The
neutron separation energy in $^{52}$Ca is a noteworthy example.  The
atomic mass table evaluation~\cite{audi2003} gives the value
$S_n=4.7$~MeV (estimated from systematics) for this nucleus.  The very
recent measurement~\cite{Dilling}, however, yields $S_n\approx 6$~MeV,
in much better agreement with our prediction of $S_n=6.59$~MeV.

\begin{table}[b]
  \begin{ruledtabular}
    \begin{tabular}{|c||l|l|l|}
      & $^{48}$Ca &$^{52}$Ca &$^{54}$Ca  \\\hline
      $E_{2^+}$(CC) & 3.58 & 2.19 &  1.89\\
      $E_{2^+}$(Exp)& 3.83 & 2.56 &  n.a. \\\hline
      $E_{4^+}/ E_{2^+}$(CC)& 1.17 & 1.80 &  2.36 \\
      $E_{4^+}/ E_{2^+}$(Exp)& 1.17 & n.a. & n.a.  \\\hline
      $S_{n}$(CC)     & 9.45 & 6.59    &  4.59   \\
      $S_{n}$(Exp)    & 9.95 & 6.0$^*$ &  4.0$^\dagger$ \\\hline
    \end{tabular}
  \end{ruledtabular}
  \caption{Excitation energies $E_{2^+}$ (in MeV) of the lowest-lying 
    $J^\pi=2^+$ states, neutron-separation energies $S_n$ (in MeV), 
    and ratio of energies $E_{4^+}/E_{2^+}$ from coupled-cluster theory (CC) 
    compared to available data (Exp) [n.a. = not available; $^*$ = from 
    Ref.~\cite{Dilling}; $^\dagger$=from atomic mass table evaluation~\cite{audi2003}]. 
    The theoretical results point to a sub-shell closure in $^{54}$Ca.}
  \label{tab1}
\end{table} 

\begin{figure}[htbp]
  \begin{center}
    \includegraphics[width=0.45\textwidth,clip=]{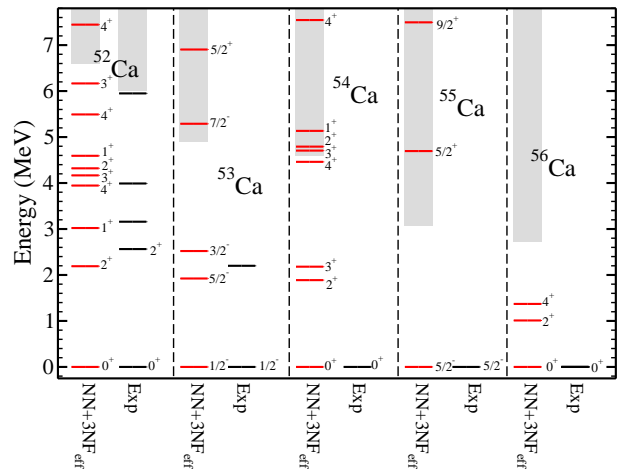}
  \end{center}
  \caption{(Color online) Theoretical excitation spectra of
    $^{52-56}$Ca compared to available data. The continuum is
    indicated as a gray area.}
  \label{fig3}
\end{figure}

To further study the evolution of shell structure we focus on spectra.
Figure~\ref{fig3} shows the computed spectra for $^{52-56}$Ca, and
compares them to available data. In $^{52,53}$Ca, our calculations
suggest spin assignments for measured levels~\cite{perrot2005}, while
we give predictions for several levels in $^{52,53,54,55,56}$Ca.  The
scattering continuum is shown as a gray band. Note that we computed
the $J^\pi=2^+$ excited state of $^{54}$Ca as a 1$p$-1$h$ excitation
of the $^{54}$Ca ground state and as $2p$ excitations of the $^{52}$Ca
ground state. In $17$ major oscillator shells we obtain
$E_{2^+}=1.80$~MeV and $E_{2^+}=1.90$~MeV, respectively. The two
methods are in good agreement with each other, and this shows the
quality of the employed methods.

In the naive shell-model picture, the order of single-particle
orbitals above the Fermi surface of $^{60}$Ca is $g_{9/2}$, $d_{5/2}$
and $s_{1/2}$.  Mean-field calculations predict that the corresponding
canonical orbitals are close in energy~\cite{meng}, yielding
deformation in the isotopes $^{60-70}$Ca~\cite{witek2012}.  The
isotope $^{58}$Ca is the heaviest isotope of calcium that has been
produced~\cite{tarasov2009}, and $^{52}$Ca is the most neutron-rich
isotope whose mass is known~\cite{Dilling}.  Nuclear energy
functionals and mass models \cite{meng,witek2012} predict the neutron
drip line to be around $^{70}$Ca.  Figure~\ref{fig3} shows that the
$J^\pi=5/2^+$ resonant excited states in $^{53,55}$Ca are lower in
energy than the $J^\pi=9/2^+$ states, and this is a deviation from the
naive shell-model picture.  There is no $J^\pi=1/2^+$ resonance in the
continuum due to absence of a centrifugal barrier.  Due to the large
$l=4$ centrifugal barrier the $J^\pi=9/2^+$ resonances can be
considered as quasi-bound states, while the $J^\pi=5/2^+$ has a
significantly larger width, see Table~\ref{tab2} for details.  Our
results shown in Fig.~\ref{fig1} yield $^{60}$Ca unbound with respect
to $^{56}$Ca.  However, we cannot rule out the existence of $^{60}$Ca
since correlations beyond triples in the coupled-cluster expansion may
play a larger role for $^{60}$Ca than for the neighbors of
$^{54}$Ca. Our computations (including the scattering continuum)
suggest that $^{61}$Ca is a very interesting nucleus. We find resonant
states $J^\pi=5/2^+$ and $J^\pi=9/2^+$ at excitation energies of 1.1
MeV and 2.2 MeV above threshold, respectively, and a virtual
$J^\pi=1/2^+$ state practically at threshold. This ordering of the
resonances is consistent with the results for
$^{53,55}$Ca. Table~\ref{tab2} summarizes our results for the
$J^\pi=5/2^+$ and $J^\pi=9/2^+$ resonances in $^{53,55,61}$Ca.  We
also find the $^{62}$Ca ground state to be very close to threshold,
unbound by about 0.2~MeV with respect to $^{60}$Ca and entirely
dominated by $(s_{1/2})^2$ configurations. In our calculations, the
correct treatment of the continuum is essential: (i) in $^{55}$Ca the
scattering continuum lowers the energy of the excited $J^\pi=5/2^+$ by
about 2~MeV when compared to a harmonic oscillator basis, (ii) in
$^{61}$Ca, the oscillator basis yields the level ordering of the naive
shell model.

\begin{table}[h]          
\begin{tabular}{|l|l|l|l|l|l|l|l|}\hline            
\multicolumn{1}{|c|}{} & \multicolumn{2}{c|}{$^{53}$Ca }  & \multicolumn{2}{c|}{$^{55}$Ca }  & \multicolumn{2}{c|}{$^{61}$Ca} \\ \hline 
 $J^\pi$    &Re[$E$] & $\Gamma$&Re[$E$] & $\Gamma$ & Re[$E$]  & $\Gamma$ \\ \hline  
$5/2^+$     & 1.99   &  1.97  &  1.63   &  1.33  &     1.14  &  0.62  \\            
$9/2^+$     & 4.75   &  0.28  &  4.43   &  0.23  &     2.19  &  0.02  \\ \hline          
\end{tabular}        
\caption{Energies of the ${5/2}^+$ and ${9/2}^+$ resonances in $^{53,55,61}$Ca. 
  ${\rm Re}[E]$ is the energy relative to the one-neutron emission threshold, 
  and the width is $\Gamma = -2{\rm Im}[E]$ (in MeV).}
\label{tab2}  
\end{table}

\begin{figure}[htbp]
  \begin{center}
    \includegraphics[width=0.45\textwidth,clip=]{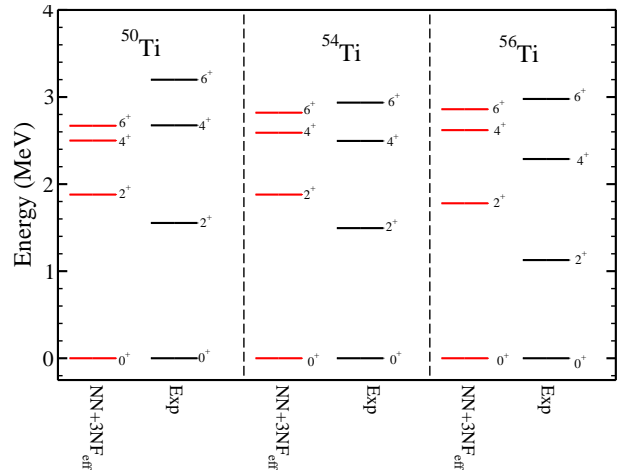}
  \end{center}
  \caption{(Color online) Theoretical excitation spectra of $^{50,54,56}$Ti compared to data.}
  \label{fig4}
\end{figure}

As a check, we also computed the lowest excited states with spin and
parity $J^\pi=2^+,4^+,6^+$ in $^{50,54,56}$Ti. These states result
from attaching two protons to the closed-core reference nuclei
$^{48,52,54}$Ca, respectively, and we employ 2$p$-0$h$ and 3$p$-1$h$
excitations in the computation.  Figure~\ref{fig4} shows a reasonably
good agreement with the data, with a maximal deviation of about
0.6~MeV for the $2^+_1$ state in $^{56}$Ti. Let us compare the
titanium isotopes to the isotopes $^{50,54,56}$Ca which might also be
computed by attaching two neutrons to $^{48,52,54}$Ca, respectively.
The analysis of the cluster amplitudes shows that the 2$p$-0$h$
cluster amplitudes account for about 60\% of the total amplitude for
the isotopes of titanium, while this number is about 72\% for the
calcium isotopes (with the remaining weight carried by 3$p$-1$h$
amplitudes).  It is insightful to compare the ratios of 3$p$-2$h$ to
2$p$-0$h$ amplitudes. These are about 0.67 in titaniums and only 0.39
in calciums. Thus, the isotopes of titanium are more correlated due to
proton-neutron interactions than the corresponding calcium
isotopes. We believe that one needs 4$p$-2$h$ amplitudes for a more
accurate description.

{\it Summary.} -- We employed interactions from chiral effective field
theory and performed coupled-cluster computations of neutron-rich
calcium isotopes. The evolution of shell structure is understood by
our computations of neutron-separation energies, low-lying
$J^\pi=2^+,4^+$ states, and excitations in the odd-mass nuclei
$^{53,55,61}$Ca.  Our results confirm that the shell closure in
$^{48}$Ca is due to the effects of three-nucleon forces, and that
$^{52}$Ca exhibits a sub-shell closure while $^{54}$Ca exhibits only a
weak sub-shell closure.  We make several predictions for spin and
parity assignments in the spectra of very neutron-rich isotopes of
calcium and predict a level ordering in $^{53,55,61}$Ca that is at
variance with the naive shell model.  Overall, we find that effects of
three-nucleon forces and the scattering continuum are essential in
understanding the evolution of shell structure towards the dripline.

\begin{acknowledgments}
  We acknowledge valuable discussions with S.~K.~Bogner, B.~A.~Brown, 
R.~J.~Furnstahl, K.~Hebeler, J.~D.~Holt, W.~Nazarewicz, and A.~Schwenk.
  This work was supported by the Office of Nuclear Physics,
  U.S.~Department of Energy (Oak Ridge National Laboratory).  This
  work was supported in part by the U.S. Department of Energy under
  Grant Nos.~DE-FG02-03ER41270 (University of Idaho),
  DE-FG02-96ER40963 (University of Tennessee), and DE-FC02-07ER41457
  (UNEDF SciDAC).  This research used computational resources of the
  National Center for Computational Sciences, the National Institute
  for Computational Sciences, and the Notur project in Norway.
\end{acknowledgments}


\begin{thebibliography}{99}

\bibitem{MJ}
M.~Mayer and J.~H.~D. Jensen, {\it Elementary Theory of
Nuclear Shell Structure}, Wiley, New York (1955).

\bibitem{sorlin2008}
O.~Sorlin and M.-G.~Porquet, 
Prog.~Part.~Nucl.~Phys.~{\bf 61}, 602 (2008).

\bibitem{otsuka2005}
T.~Otsuka {\it et al.}, 
\prl {\bf 95}, 232502 (2005)

\bibitem{zuker2003}
A.~P.~Zuker, 
\prl {\bf 90}, 042502 (2003). 

\bibitem{otsuka2010} 
T.~Otsuka  {\it et al.}, Phys.~Rev.~Lett.~{\bf 105}, 032501 (2010).

\bibitem{jdholt2011b}
J. D. Holt and A. Schwenk, arxiv:1108:2680 (2011).

\bibitem{hagen2012} 
G.~Hagen {\it et al.}, 
Phys.~Rev.~Lett.~{\bf 108}, in press  (2012) and arXiv:1202.2839 (2012).

\bibitem{jdholt2011}
J.~D.~Holt {\it et al.}, 
arXiv:1009.5984v2 (2011).

\bibitem{huck1985} 
A.~Huck {\it et al.}, 
 Phys. Rev. C {\bf 31}, 2226 (1985). 

\bibitem{gade2006}
A.~Gade {\it et al.}, 
Phys. Rev. C {\bf 74}, 021302 (2006).


\bibitem{janssens2002} 
  R.~V.~F. Janssens {\it et al.},
Phys. Lett. B {\bf 546}, 55 (2002).


\bibitem{prisciandaro2001} 
J.~I.~Prisciandaro {\it et al.}, 
Phys. Lett. B {\bf 510}, 17 (2001).

\bibitem{marginean2006}
N.~Marginean {\it et al.}, Phys. Lett. B {\bf 633}, 696 (2006)
  
\bibitem{liddick2004}
S.~N.~Liddick {\it et al.},
Phys. Rev. Lett. {\bf 92}, 072502 (2004).

\bibitem{dinca2005}
D.-C.~Dinca {\it et al.},
Phys. Rev. C {\bf 71}, 041302 (2005).

\bibitem{rejmund2007}
M.~Rejmund {\it et al.}, 
\prc {\bf 76}, 021304 (2007). 


\bibitem{gxpf1} 
M.~Honma {\it et al.}, 
Phys.~Rev.~C~{\bf 65}, 061301 (2002).

\bibitem{kb3} E.~Caurier {\em et al.}, Rev.~Mod.~Phys.~{\bf 77}, 427 (2005).

\bibitem{luigi2009}
L.~Coraggio {\it et al.}, 
\prc {\bf 80}, 044311 (2009).

\bibitem{jacek2007} J.~Dobaczewski {\em et al.}, Prog.~Part.~Nucl.
Phys. {\bf 59}, 432 (2007).

\bibitem{entem2003} D.~R.~Entem and R.~Machleidt, Phys.~Rev.~C {\bf
    68}, 041001(R) (2003).

\bibitem{machleidt2011} R.~Machleidt and D.~R.~Entem, Phys.~Rep.~{\bf
    503}, 1 (2011).


\bibitem{ccm}
F.\ Coester, Nucl.\ Phys.\ {\bf 7}, 421 (1958); 
F.\ Coester and H.\ K{\" u}mmel, Nucl.\ Phys.\ {\bf 17}, 477 (1960);
J.\ {{\v C}{\'\i}{\v z}ek}, J.\ Chem.\ Phys.\ {\bf 45}, 4256 (1966); 
J.\ {{\v C}{\'\i}{\v z}ek}, Adv.\ Chem.\ Phys.\ {\bf 14}, 35 (1969);
H.\ K{\" u}mmel, K.H.\ L{\" u}hrmann, and J.G.\ Zabolitzky, 
Phys.\ Rep.\ {\bf 36}, 1 (1978).

\bibitem{bartlett07}
R.~J.~Bartlett and M.~Musia{\l}, Rev.~Mod.~Phys. {\bf 79}, 291 (2007).



\bibitem{holt2009} J.~W.~Holt, N.~Kaiser, and W.~Weise, 
\prc {\bf 79}, 054331 (2009);
{\it ibid.} Phys.~Rev.~C {\bf 81}, 024002 (2010).

\bibitem{hebeler2011} 
K.~Hebeler and A.~Schwenk, Phys.~Rev.~C {\bf 82}, 014314 (2010); 
K.~Hebeler  {\it et al.}, 
Phys.~Rev.~C {\bf 83}, 031301 (2011).

\bibitem{hagen2007}
G.~Hagen {\it et al.}, 
\prc {\bf 76}, 034302 (2007).


\bibitem{gazit}
D.~Gazit, S.~Quaglioni, and P.~Navr{\'a}til, 
Phys. Rev. Lett. {\bf 103}, 102502 (2009).




\bibitem{hagen2008}
G.~Hagen {\it et al.}, 
Phys. Rev. Lett. {\bf 101}, 092502 (2008).

\bibitem{hagen2010a}G.~Hagen  {\it et al.}, 
Phys.~Rev.~C {\bf 82}, 034330 (2010).

\bibitem{taube2008} A.~D.~Taube and R.~J.~Bartlett, J.~Chem.~Phys. {\bf 128}, 044110 (2008);
{\it ibid.} {\bf 128}, 044111 (2008).


\bibitem{michel}
N.~Michel  {\it et al.}, 
J.~Phys.~G {\bf 36}, 013101 (2009).

\bibitem{hagen2006}
G.~Hagen and J.~S.~Vaagen, 
Phys. Rev. C. {\bf 73}, 064307 (2006).

\bibitem{jansen2011} G.~R.~Jansen {\it et al.}, 
Phys.~Rev.~C.~{\bf 83}, 054306 (2011).

\bibitem{Dilling} A.~Lapierre {\it et al.}, Phys.~Rev.~C {\bf 85},
  024317 (2012); A.~T.~Gallant {\it et al.}, arXiv:1204.1987 (2012).

\bibitem{audi2003}
G.~Audi, A.~H.~Wapstra, and C.~Thibault,
Nucl. Phys. A {\bf 729}, 337 (2003).

\bibitem{perrot2005} 
F.~Perrot {\it et al.}, \prc {\bf 74}, 014313 (2005).

\bibitem{meng}
J.~Meng {\it et al.}, \prc {\bf 65}, 041302 (2002); 
S.~A.~Fayans, S.~V.~Tolokonnikov, and D.~Zawischa, Phys.~Lett.~B {\bf 491}, 245 (2000).

\bibitem{witek2012}  W.~Nazarewicz {\em et al.}, Phys.~Rev.~C {\bf 53}, 740 (1996);
J.~Erler {\em et al.}, to be published (2012).

\bibitem{tarasov2009}
O.~B.~Tarasov {\it et al.}, Phys.~Rev.~Lett.~{\bf 102}, 142501 (2009); O.~B.~Tarasov, private communication (2012).


\end{thebibliography}
\end{document}